\documentclass{article}
\usepackage{amsmath}
\usepackage{amssymb}
\usepackage[Gray,squaren,thinqspace,thinspace]{SIunits}
\usepackage{graphicx}
\usepackage[small,bf]{caption}
\usepackage{a4wide}
\usepackage{bbm}
\DeclareMathOperator{\tr}{tr}
\newcommand{\lms}{\Lambda_{\overline{\mbox{\tiny{MS}}}}}

\begin{document}

\title{A two-component picture of the $\langle A_\mu^2\rangle$ condensate with instantons}
\author{David Vercauteren\thanks{E-mail: David.Vercauteren@UGent.be}, Henri Verschelde\thanks{E-mail: Henri.Verschelde@UGent.be}\\\\
\textit{\tiny Ghent University, Department of Physics and Astronomy, Kr\ij gslaan 281-S9, B-9000 Gent, Belgium}}
\date{}

\maketitle

\begin{abstract}
We study the interplay between the $\langle A_\mu^2\rangle$ condensate and instantons in non-Abelian gauge theory. Therefore we use the formalism of Local Composite Operators, with which the vacuum expectation value of this condensate can be analytically computed. We first use the dilute gas approximation and partially solve the infrared problem of instanton physics. In order to find quantitative results, however, we turn to an instanton liquid model, where we find how the different contributions to the condensate add up.
\end{abstract}

\section{Introduction}
The dimension 2 gluon condensate $\langle A_\mu^2\rangle$ in pure Yang--Mills theory has been proposed in \cite{Gubarev:2000eu,Gubarev:2000nz}, and it has been investigated in different ways since then \cite{lcos, Dudal:2002pq, Dudal:2003vv, Vercauteren:2007gx, Dudal:2005na,Boucaud:2001st, Furui:2005he, Gubarev:2005it, Browne:2003uv, Andreev:2006vy, RuizArriola:2006gq, Chernodub:2008kf}.

In \cite{lcos} an analytical framework for studying this condensate has been developed, based on work carried out in the Gross--Neveu model \cite{verscheldegn}. Different problems had to be overcome. First of all there is the gauge invariance of this condensate. In order to make the operator $A_\mu^2$ gauge invariant, one can take the minimum of its integral over the gauge orbit. Since $\int d^dx\, A_\mu^U A_\mu^U$, with $U\in SU(N)$, is positive, this minimum will always exist. In a general gauge, however, the minimum is a highly non-local and thus hard to handle expression of the gauge field. A minimum is however reached in the Landau gauge ($\partial_\mu A_\mu=0$), such that working in this gauge reduces the operator to a local expression\footnote{We ignore the Gribov problem here, see also \cite{Dudal:2005na}.}. Secondly adding a source $J$, coupled to $A_\mu^2$, makes the theory non-renormalizable at the quantum level. To solve this, a term quadratic in the source must be added, which in turn spoils the energy interpretation of the effective action. One way around this is to perform the Legendre inversion, but this is rather cumbersome, especially with a general, space-time dependent source. One can also use a Hubbard-Stratonovich transform, which introduces an auxiliary field (whose interpretation is just the condensate) and eliminates the term quadratic in the source. Details can be found in \cite{lcos}. The result was that the Yang--Mills vacuum favors a finite value for the expectation value of $A_\mu^2$. The precise renormalization details of the procedure proposed in \cite{lcos} were given in \cite{Dudal:2002pq}. We review this formalism in Section \ref{LCO}.

Instantons play an important role in the QCD vacuum and have a large influence in many infrared properties (see \cite{Schafer:1996wv} for a review). As such it is an interesting question what their connection with the dimension two condensate is. A first study in this direction has been done on the lattice by Boucaud \emph{et al.} \cite{Boucaud:2002nc,Boucaud:2002wy}, and a rather large instanton contribution to the condensate has been found, which shows some agreement with the results from an OPE approach to the gluon propagator from \cite{Boucaud:2001st}. However, the condensate may get separate contributions from other sources, as for example the non-perturbative high-energy fluctuations leading to the condensate found in \cite{lcos}. The opposite viewpoint is just as interesting: what is the influence of an effective gluon mass on the instanton ensemble? In 't~Hooft's seminal paper he found that, in a Higgs model, a gauge boson mass stabilizes the instanton gas\cite{thooft}.

Some subtle points are to be resolved before a full treatment can be given. These are discussed in Section \ref{instantonakwadraat}. Then, Section \ref{oneloop} is devoted to the computation of the one-loop effective action, for which we use the strategy developed by Dunne \emph{et al.}\cite{dunneprl,dunnelang}. Finally, Section \ref{conclusions} concludes this Letter.

\section{$\langle A_\mu^2\rangle$ and instantons} \label{LCO}
In this section we will review the LCO formalism as proposed in \cite{lcos} and modified to use it with a background field.

As a first step the gauge is fixed using the Landau condition, i.e. the linear covariant gauge $\partial_\mu A_\mu = 0$ with $\xi \rightarrow 0$. Then, a term
\begin{equation}
\frac12 J A_\mu^2
\end{equation}
is added to the Lagrangian density. Here $J$ is the source which will be used to compute $\langle A_\mu^2\rangle$. As it stands, the theory is not renormalizable. To correct this, a new term
\begin{equation}
- \frac12 \zeta J^2
\end{equation}
has to be added. Here $\zeta$ is a new coupling constant which will have to be determined as a function of the parameters in the original theory. This Lagrangian is now multiplicatively renormalizable, as shown in \cite{brst} using a BRST analysis.

As we want to work with an instanton as a background field, it is more appropriate to use the Landau background gauge \cite{Abbott:1981ke} $\mathcal D_\mu[\hat A] A_\mu = 0$ instead of the usual Landau gauge prescription $\partial_\mu A_\mu = 0$. Here $\hat A_\mu$ is the background field. In order to do so, some alterations are in order. A BRST analysis (for BRST in the background gauge, see for example \cite{brstbackground}) shows that, in order for the LCO formalism to stay renormalizable, the condensate $A_\mu^2$ must be replaced by
\begin{equation}
(A_\mu - \hat A_\mu)^2 = \mathcal A_\mu^2
\end{equation}
with $A_\mu$ the total gauge field and $\mathcal A_\mu$ the quantum fluctuations, $A_\mu = \mathcal A_\mu + \hat A_\mu$.

In order for this formalism to work, some creases have to be ironed out. As a first point, we have introduced a new parameter, $\zeta$, creating a problem of uniqueness. However, it is possible to choose $\zeta$ to be a unique meromorphic function of $g^2$ based on the renormalization group equations. In \cite{lcos} there was found, using the $\lms$ scheme in $d=4-\epsilon$ dimensions and without any background field (up to one-loop order and with $N_c$ the number of colors):
\begin{subequations}
\begin{gather}
\zeta = \frac{9}{13}\frac{N_c^2-1}{N_c} \frac1{g^2} + \frac{N_c^2-1}{16\pi^2}\frac{161}{52} + \mathcal O(g^2) \\
Z_\zeta = 1 - \frac{g^2N_c}{16\pi^2} \frac{13}{3\epsilon} + \mathcal O(g^2) \\
Z_2 = 1 - \frac{N_cg^2}{16\pi^2} \frac3{2\epsilon} + \mathcal O(g^2)
\end{gather}
\end{subequations}
where $Z_\zeta$ and $Z_2$ are the constants renormalizating $\zeta J^2$ and $J\mathcal A_\mu^2$ respectively. For dimensional reasons, working in the background gauge will change nothing to the expressions for $\zeta$ and the renormalization constants.

Secondly the presence of the $J^2$ term spoils an energy interpretation for the effective potential. One way around this is to perform the Legendre inversion, but this is rather cumbersome, especially so with a general, space-time dependent source. A more elegant way out applies a Hubbard--Stratonovich transformation by inserting unity into the path integral:
\begin{equation}
1 = \mathcal N \int[\mathcal D\sigma] \exp-\frac1{2\zeta} \int \left(\frac\sigma g + \frac12\mathcal A_\mu^2 - \zeta J\right)^2 d^4x
\end{equation}
with $\mathcal N$ an irrelevant constant. This eliminates the $\frac12J\mathcal A_\mu^2$ and $\zeta J^2$ terms from the Lagrangian and introduces a new field $\sigma$. The result is:
\begin{equation}
e^{-W(J)} = \int[\mathcal D\mathcal A_\mu][\mathcal D\sigma] \exp - \int\left(\mathcal L_\text{YM} [\mathcal A_\mu,\hat A_\mu,c,\bar c] + \mathcal L_\text{LCO} [\mathcal A_\mu,\hat A_\mu,\sigma] - \frac\sigma g J\right) d^4x \;.
\end{equation}
Here $\mathcal L_\text{YM}$ is the well-known Yang--Mills Lagrangian with Faddeev--Popov ghosts, fixed in the Landau background gauge, and
\begin{equation}
\mathcal L_\text{LCO} [\mathcal A_\mu,\sigma] = \frac{\sigma^2}{2 g^2\zeta} + \frac{\sigma \mathcal A_\mu^2}{2g\zeta} + \frac{(\mathcal A_\mu^2)^2}{8\zeta} \;.
\end{equation}
Now $J$ acts as a linear source for the $\sigma$ field, so that we can straightforwardly compute the effective action $\Gamma(\sigma)$ using the above expressions.

If we compare our new Lagrangian to the original expression, we find that the expectation value of $\sigma$ corresponds to the expectation value of the composite operator
\begin{equation}
\langle\sigma\rangle = -g\left\langle\frac12\mathcal A_\mu^2 - \zeta J\right\rangle \;.
\end{equation}
In the limit $J\rightarrow0$ this operator corresponds (up to a multiplicative factor) to $\mathcal A_\mu^2$. We can also read off the effective gluon mass in the lowest order:
\begin{equation} \label{meff}
m^2 = \frac\sigma{g\zeta} = \frac{N_c}{N_c^2-1}\frac{13}{9} g\sigma \;.
\end{equation}

\section{Instantons and $\langle A_\mu^2\rangle$} \label{instantonakwadraat}
Let us first look into whether the condensate $\langle A_\mu^2\rangle$ can stabilize the instanton ensemble in the LCO formalism, as, if successful, it would minimize the amount of hand-waving necessary to compute the action. First we have the question of which gauge to choose. All instanton calculations are done in background gauges, as analytic computations in non-background gauges are quite impossible. The LCO formalism does not give classical fields a mass in the Landau background gauge, however. In the electroweak theory considered by 't~Hooft in \cite{thooft} it is exactly this classical mass which suppresses large instantons by the simple fact that large instantons are no solutions to the massive field equations anymore, while small instantons can still be considered approximate solutions.

If we want to have a mass already at the classical level, it is necessary to work in the non-background Landau gauge. Although the computations cannot be carried through in this gauge, it still possible to find the qualitative form of the result. In order to circumvent the question of which background to take for the $\sigma$ field\footnote{Allowing $\sigma$ to obey its own \emph{classical} field equations does not lead to non-trivial results.} it is more opportune to start before the point where the Hubbard--Stratonovich transformation is introduced.

We start from
\begin{equation}
-\frac12\langle A_\mu^2\rangle = \left. \frac\delta{\delta J} \ln \int [dA_\mu] e^{-S-\frac12JA_\mu^2+\frac\zeta2J^2} \right|_{J=0} \;.
\end{equation}
As the source is small, instantons will be approximate solutions. Eventually, we can correct the instanton using the valley method\cite{Affleck:1980mp}, but this turns out not to give more insight. At the classical level, the action of the instanton is now
\begin{equation}
S + \frac12JA_\mu^2 = \frac{8\pi^2}{g^2} + \frac{6\pi^2}{g^2}J\rho^2 + \cdots \;,
\end{equation}
where the dots stand for contributions from corrections to the instanton solution. From renormalization group arguments, we can now write down the general form of the one-loop result:
\begin{equation}
W[J] = W^{0I}[J] - \int_0^\infty \frac{d\rho}{\rho^5} \exp\left(-\frac{8\pi^2}{g^2} - \frac{6\pi^2}{g^2}J\rho^2 + \frac{11}3\ln(\mu^2\rho^2) + f_1(J\rho^2) + \cdots\right) \;,
\end{equation}
where the dilute instanton gas approximation has been used, giving an exponential of the instanton contribution. Here, $W^{0I}$ stands for the zero-instanton result, and $f_1$ is an unknown function which gives the quantum corrections. A factor of the space-time volume has been left out. For finite $J$, the integral over the instanton size $\rho$ is now convergent and can be done:
\begin{equation}
W[J] = W^{0I}[J] - g^{10/3}\mu^{22/3}J^{-5/3} e^{-\frac{8\pi^2}{g^2}} f_2(g^2) \;,
\end{equation}
where $f_2$ is a new unknown function. Mark that the limit $J\rightarrow0$ gives 't~Hooft's divergent result again. Doing the Legendre inversion yields
\begin{equation}
\Gamma[\sigma] = \Gamma^{0I}[\sigma] - g^{10/3}\mu^{22/3}\sigma^{-5/3} e^{-\frac{8\pi^2}{g^2}} f_3(g^2) \;,
\end{equation}
where $f_3$ is yet another unknown function, and $\Gamma^{0I}$ is the zero-instanton result. If the coupling is sufficiently small, the instanton correction can be ignored and the zero-instanton result is recovered. The instanton term can then be considered as a small perturbation, slightly shifting the value of the condensate. However, no matter how small the coupling, the second term will always diverge for sufficiently small $\sigma$, and so the effective action will be unbounded from below\footnote{It is easy to see that $f_3(g^2)$ must be positive, at least for small $g^2$.}. This is of course related to the infrared divergence found in the case without condensate.

The conclusion is that two problems can be identified. First there is the resilience of the infrared divergence. One could say this is due to the strength of the LCO formalism ---the gluon mass is left free in order to determine it by the gap equation, which allows the possibility for the mass to be zero, which again allows instantons to proliferate and to so destabilize the action. This can be solved invoking only a little hand-waving: when $\sigma$ is small the dilute instanton gas approximation is not valid, and so this part of the result must be thrown away. The final conclusion is that instantons slightly shift the value of $\langle A_\mu^2\rangle$.

This leaves a second problem: one would expect each instanton to give a contribution of $12\pi^2\rho^2/g^2$ to the condensate already at the classical level. This does not happen, which is due to the way the problem has been approached. The dilute instanton gas approximation starts from the one-instanton contribution and exponentiates it to give a gas. The contribution of one instanton to the condensate is negligible ---it is finite, while the total condensate is proportional to the space-time volume--- and so it drops out.

In the background gauge this last problem is readily solved: the classical and quantum mechanical contributions are neatly separated from the start. Furthermore it turns out that the computations can all be done, which allows for a quantitative result to be given as well. Only the infrared divergence still remains as a problem, but, as some hand-waving is necessary anyway, one of the many instanton liquid models can be used to cure this. This is the subject of the following section.

\section{Computing the one-loop determinant} \label{oneloop}
In the background gauge, $\mathcal L_\text{LCO} [\mathcal A_\mu,\hat A_\mu,\sigma]$ does not change the classical field equations for $A_\mu$, as at the classical level we have that $A_\mu\equiv\hat A_\mu$, making $\mathcal L_\text{LCO}$ vanish. This means that the instanton will not be modified as in the non-background gauge in the previous section or as in electroweak theory ---in our case a vacuum expectation value for $\sigma$ will only give a mass to the quantum fluctuations, not to the classical part of $A_\mu$.

The computation of the one-loop quantum corrections to the action of massive fields in an instanton background is a non-trivial feat. Recently, Dunne \emph{et al.} have developed a strategy leading to an exact albeit numerical result\cite{dunneprl,dunnelang}. We give a short overview of the necessary steps as applied to spin and isospin 1 fields. More details can be found in \cite{dunnelang}.

We expand around a constant value for $\sigma$ and around a one-instanton configuration for $A_\mu^a$. The quantum fluctuation in $\sigma$ can be immediately integrated out, and we find that up to one-loop order:
\begin{multline}
V_\text{eff} = \frac{8\pi^2}{g^2} + V \frac{\sigma^2}{2 g^2\zeta} - \log\det(-\mathcal{D}^2) \\
+ \frac12 \log\det\left(- g_{\mu\nu} \mathcal{D}^2_{ab} + \left(1-\frac1\xi\right) (\mathcal{D}_\mu\mathcal{D}_\nu)^{ab} + 2g\epsilon^{abc} F_{\mu\nu}^c + \frac\sigma{g\zeta} g_{\mu\nu}\delta^{ab}\right)
\end{multline}
where all covariant derivatives contain only the instanton background, where the limit $\xi\rightarrow0$ for the Landau gauge is implied, and with $V$ the volume of space-time.

The $\log\det$ of the gluon propagator can be simplified as in 't~Hooft's original paper\cite{thooft}\footnote{'t~Hooft does not mention spin elimination for gluons, only for fermions, but the procedure is essentially the same.}. The presence of a mass combined with the Landau gauge instead of the Feynman gauge complicate matters slightly, however. First, suppose we have a function obeying
\begin{equation}
-\mathcal D^2\psi^a = \lambda\psi^a \,
\end{equation}
then one can show that
\begin{equation}
\Delta_{\mu\nu}^{ab} \mathcal D_\nu\psi^b = \left(\frac\lambda\xi + \frac\sigma{g\zeta}\right) \mathcal D_\mu\psi^a
\end{equation}
where $\Delta_{\mu\nu}^{ab}$ is the gluon propagator in a one-instanton background. In order to find this result, one has to make use of the classical field equations for $A_\mu$. We see that, in the limit $\xi\rightarrow0$, the functions $\mathcal D_\mu\psi^a$ will become massless and they will give a contribution of $\frac12\log\det(-\mathcal D^2) + \frac12\tr\ln\xi$ to the effective action, and they will cancel half of the ghost contribution. A second contribution comes from the functions $\bar\eta^i_{\mu\nu}\mathcal D_\nu\psi^a$ with $i=1,2,3$. Using the properties of the 't~Hooft symbols and the explicit form of the instanton, it is straightforward to show that these functions obey $\mathcal D_\mu A_\mu^a = 0$. For these functions, we get
\begin{equation}
\Delta_{\mu\nu}^{ab}\bar\eta^i_{\nu\lambda}\mathcal D_\lambda\psi^a = \left(\lambda + \frac\sigma{g\zeta}\right) \bar\eta^i_{\mu\nu}\mathcal D_\nu\psi^a \;,
\end{equation}
meaning they will contribute $\frac32\log\det(-\mathcal D^2+\sigma/g\zeta)$ to the effective action. This leaves us with
\begin{equation}
V_\text{eff} = \frac{8\pi^2}{g^2} + V\frac{\sigma^2}{2 g^2\zeta} - \frac12\log\det(-\mathcal{D}^2) + \frac32\log\det\left(-\mathcal D^2 + \frac\sigma{g\zeta}\right) \;.
\end{equation}

In the above arguments we have ignored the existence of zero modes, which cannot be written as covariant derivatives of some Lorentz-scalar function. So they have to be considered separately. Due to the classical action being the unmodified Yang--Mills action, one would naively expect these modes to remain zero modes. However, going through the computations uncovers that they get a mass $\sigma/g\zeta$. This is due to the perturbative approximation. Properly including all the interactions between $\sigma$ and the gluon field to all orders will make the zero modes massless again, and we will treat them as such here. Using the action without the Hubbard--Stratonovich transformation and with a source directly coupled to $A_\mu^2$ shows that this is indeed the right course.

In order to compute the functional determinants, it is convenient to split off the zero-instanton contributions:
\begin{equation} \label{veff}
V_\text{eff} = V V_\text{eff}^{0I} + \frac{8\pi^2}{g^2} - \frac12\log\det\left(\frac{-\mathcal{D}^2}{-\partial^2}\right) + \frac32\log\det\left(\frac{-\mathcal D^2 + \frac\sigma{g\zeta}}{-\partial^2 + \frac\sigma{g\zeta}}\right) \;.
\end{equation}
In the above equation,
\begin{equation}
V V_\text{eff}^{0I} = V\frac{\sigma^2}{2 g^2\zeta} -\frac12\log\det(-\partial^2) + \frac32\log\det\left(-\partial^2+\frac\sigma{g\zeta}\right)
\end{equation}
is the zero-instanton action and $V$ is the space-time volume. 't~Hooft already computed the first functional determinant in \eqref{veff}, finding
\begin{equation}
\log\det\left(\frac{-\mathcal{D}^2}{-\partial^2}\right) = \frac13\left(\frac2\epsilon+\ln\rho^2\bar\mu^2\right) - 8\zeta'(-1) - \frac{10}9 + \frac13\ln2
\end{equation}
where $\epsilon = 4-d$ with $d$ the number of dimensions in dimensional regularization and $\bar\mu$ is the scale set by going to the $\lms$ scheme. For the second functional determinant, the work by Dunne and collaborators is to be followed.

As a first step, the operators under consideration are separated in a radial part and an angular and isospin part. The angular and isospin quantum numbers couple according to the usual spin-orbit coupling mechanism, and we can write:
\begin{equation}
\log\det\left(\frac{-\mathcal D^2 + \frac\sigma{g\zeta}}{-\partial^2 + \frac\sigma{g\zeta}}\right) = \sum_{j=0,\frac12,1\ldots}^{+\infty} \sum_{l=|j-1|}^{j+1} (2l+1)(2j+1) \log\det\left(\frac{-\mathcal D^2_{l,j} + \frac\sigma{g\zeta}}{-\partial^2_l + \frac\sigma{g\zeta}}\right)
\end{equation}
where the subscripts $l,j$ indicate that we take the part of the operators working in the sector with rotational quantum numbers $l$ and $j$. (The operator $-\partial^2$ does not have a $j$ dependence, so that we can leave out this index.) Now the operators in both numerator and denominator are one-dimensional and an old trick relating the functional determinant of an operator $\hat{\mathcal O}$ to the asymptotic value of a function obeying $\hat{\mathcal O} f = 0$ can be used. In our case we define
\begin{subequations} \begin{gather}
\left(-\mathcal D_{l,j}^2 + \frac\sigma{g\zeta}\right) \psi_{l,j}(r) = 0 \;,\qquad \psi_{l,j}(r) \mathop=_{r\rightarrow0} r^{2l} \;, \\
\left(-\partial_l^2 + \frac\sigma{g\zeta}\right) \psi_l^0(r) = 0 \;, \qquad \psi_l^0(r) \mathop=_{r\rightarrow0} r^{2l} \;.
\end{gather} \end{subequations}
Then we have that:
\begin{equation}
\log\det\left(\frac{-\mathcal D^2_{l,j} + \frac\sigma{g\zeta}}{-\partial^2_l + \frac\sigma{g\zeta}}\right) = \lim_{r\rightarrow\infty} \log\frac{\psi_{l,j}(r)}{\psi_l^0(r)} \;.
\end{equation}
The functions $\psi_{l,j}(r)$ and $\psi_l^0(r)$ can be found numerically. In \cite{dunneprl,dunnelang} there is explained how to find a differential equation for the logarithm of the determinant itself, which is numerically more stable, and also how the convergence of the integration can be increased.

Now it remains to regularize the sum over the quantum numbers $l$ and $j$. This sum is, of course, divergent. A first step to control this is to write the sum as:
\begin{equation}
\sum_{j=0,\frac12,1\ldots}^{+\infty} \sum_{l=|j-1|}^{j+1} f_{l,j} = \sum_{l=0,\frac12,1\ldots}^{+\infty} (f_{l,l+1} + f_{l+\frac12,l+\frac12} + f_{l+1,l})
\end{equation}
with $f_{l,j}$ our summand. This sum is now much less divergent than the original one, albeit still not finite. In order to find a finite result, the theory has to be renormalized. Therefore we introduce a Pauli--Villars regulator. If we take a certain cut-off $l=L$ in our sum, we can separate it into two parts: one with $l\leq L$, where the Pauli--Villars regulator can be taken to infinity and which can be computed numerically to give a finite result, and one part with $l>L$, which has to be computed analytically and which we will use to subtract the divergences from the numerically determined sum.

\begin{figure}\begin{center}
\includegraphics[width=.5\textwidth]{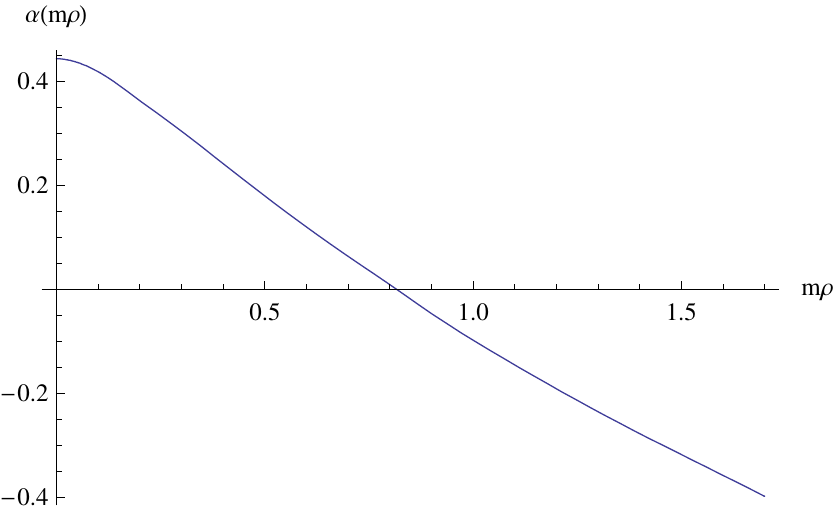}
\caption{The function $\alpha(m\rho)$ found from the computations in equation \eqref{grotevergelijking}. \label{alpha}}
\end{center}\end{figure}

This analytic computation can be done in a WKB expansion. In the limit of high $L$, only the first two orders in the WKB expansion contribute. This computation has been done by Dunne \emph{et al.} for particles in the fundamental representation of the gauge group, and the procedure can be straightforwardly applied to adjoint particles. Finally we find in dimensional regularization:
\begin{multline} \label{grotevergelijking}
\log\det\left(\frac{-\mathcal D^2 + \frac\sigma{g\zeta}}{-\partial^2 + \frac\sigma{g\zeta}}\right) = \frac13 \left(\frac2\epsilon+\ln\rho^2\bar\mu^2\right) +  \lim_{L\rightarrow\infty} \Bigg( \sum_{l=0,\frac12,1\ldots}^{L} \Gamma^S_l(\rho^2\sigma/g\zeta) + 8L^2 + 20L \\
- \ln L\left(\frac23+\frac{2\rho^2\sigma}{g\zeta}\right) + \frac{83}9 - \frac43\ln2 + \frac{\rho^2\sigma}{g\zeta} \left(2-4\ln2+\ln\frac{\rho^2\sigma}{g\zeta}\right) \Bigg)
\end{multline}
where $\Gamma_l(\rho^2\sigma/g\zeta)$ is the result from the numerical computation with quantum number $l$. Practically, taking $L\approx50$ gives acceptable results. The function defined by the limit of the expression between brackets is plotted in Fig. \ref{alpha}.

Putting everything together and working in the dilute gas approximation, which sums all contributions from all numbers of instantons into an exponential, we find for the action density:
\begin{multline} \label{effact}
\frac1V V_\text{eff}(m^2) = \frac{27}{26} \frac{m^4}{2g^2} + \frac94 \frac{m^4}{(4\pi)^2} \left(-\frac56-\frac{161}{78}+\ln\frac{m^2}{\bar\mu^2}\right) \\
+ \frac{2^{10}\pi^6}{g^8} \int_0^\infty \frac{d\rho}{\rho^5} \exp\left(-\frac{8\pi^2}{g^2} + \frac{11}3 \ln\bar\mu^2\rho^2 - \frac32\alpha(m\rho) + \frac12\alpha(0)\right) \;,
\end{multline}
where $m$ is the effective gluon mass defined in \eqref{meff}, $\alpha(m\rho)$ is the function computed numerically and shown in Fig. \ref{alpha}, and $\alpha(0) = - 8\zeta'(-1) - 10/9 + 1/3 \ln2$. The integration over the instanton size $\rho$ is divergent, as in the massless case. One might naively expect a gluon mass to cure this divergence, as happens in electroweak theory, but here the mass only enters in the quantum correction and does not operate at the classical level. Therefore, the integral is still divergent, and the contribution from $\alpha(m\rho)$ makes this even worse than when $m=0$.

In order to extract meaningful results from the effective action \eqref{effact}, the integral has to be given a finite value in some way. The easiest way out is to add an infrared cut-off $\rho_c$ as the upper bound of the integral, but this violates the scaling Ward identities\cite{Hutter:1995sc}. Several improvements have been suggested, usually involving interactions between the instantons. For our purpose, however, it suffices to take a phenomenological approach: we suppose the infrared divergence is somehow cured, and we work in an instanton liquid with certain values for the density $n$ and average radius $\rho$. This modifies the effective action to
\begin{multline}
\frac1V V_\text{eff}(m^2,n,\rho) = \frac{27}{26} \frac{m^4}{2g^2} + \frac94 \frac{m^4}{(4\pi)^2} \left(-\frac56-\frac{161}{78}+\ln\frac{m^2}{\bar\mu^2}\right) \\
- n \exp\left(-\frac{8\pi^2}{g^2} + \frac{11}3 \ln\bar\mu^2\rho^2 - \frac32\alpha(m\rho) + \frac12\alpha(0)\right) \;.
\end{multline}
Phenomenological values for $n$ and $\rho$ found on the lattice are\cite{Schafer:1996wv}
\begin{equation}
n \approx \unit{1}{\femto\meter^{-4}} \approx (\unit{0.6}{\lms})^4 \;, \qquad \rho \approx \unit{\frac13}{\femto\meter} \approx (\unit{1.8}{\lms})^{-1} \;,
\end{equation}
where $\lms = \unit{330}{\mega\electronvolt}$ in SU(2).

Taking the scale $\bar\mu^2$ at the value of $m^2$ in the global minimum of the action, we find that the instantons are much suppressed by the relative smallness of the coupling $g^2$. The non-perturbative minimum is still at $m\approx\unit{2.05}{\lms}$, as in the case without instantons. Now, however, we cannot say that $\langle \frac12g^2A_\mu^2\rangle = -\frac{27}{26}m^2 = -\unit{4.36}{\lms^2}$ in SU(2), since the instanton contribution to the condensate has to be included. As in \cite{Boucaud:2002nc,Boucaud:2002wy}\footnote{Mark that the authors of \cite{Boucaud:2002nc,Boucaud:2002wy} used a different convention for the gauge fields, and their $A_\mu^2$ corresponds to $g^2A_\mu^2$ here.} each instanton gives a contribution of $12\pi^2\rho^2$, resulting in
\begin{equation} \label{instantontotaal}
\langle g^2A_\mu^2\rangle_\text{tot} = -(\unit{2.0}{\lms})^2 = -\unit{0.42}{\giga\electronvolt^2} \;.
\end{equation}
This value depends strongly on the instanton liquid parameters plugged into the model. It is negative but close to zero because the instanton and quantum contributions are similar in magnitude but opposite in sign, and the quantum corrections have slightly larger absolute value.

\section{Conclusions} \label{conclusions}
A first conclusion arrived at in this Letter is that we have not been able to solve the infrared problem plaguing instanton physics by adding an effective gluon mass coming from the dimension two condensate. As the gluon mass must be determined from its gap equations, this leaves open the possibility of it being zero, which gives instantons the possibility to cause the infrared divergence. The amount of hand-waving necessary to stabilize the vacuum is less than without the condensate (one only has to state that the mass will be sufficiently high and the divergence is swept under the rug), but the state of affairs is not yet very satisfying.

The second main conclusion of this Letter is that, when working in the Landau background gauge, the LCO formalism gives a separate contribution to $\langle A_\mu^2\rangle$, which lowers the contributions coming from the instantons themselves.

\section*{Acknowledgements}
This work is supported financially by the ``Special Research Fund'' of Ghent University.

\end{document}